\documentclass[aps, prd, superscriptaddress, showpacs, floatfix, twocolumn, nofootinbib, reprint, preprintnumbers, longbibliography]{revtex4-2}

\usepackage{bm, amsmath, amssymb, relsize, amsfonts, mathrsfs, multirow, braket, siunitx, color, booktabs, arydshln, mathtools, calligra}

\usepackage{graphicx}

\DeclareSIUnit \parsec {pc}

\usepackage[dvipsnames]{xcolor}
\usepackage[unicode]{hyperref}
\hypersetup{
  colorlinks=true,
  citecolor=cyan,
  linkcolor=red,
  urlcolor=cyan
}

\usepackage{orcidlink}

\newcommand{\done}[1]{{\color{}}}

\newcommand{\norm}[1]{\left\lVert#1\right\rVert}

\DeclareMathAlphabet{\mathcalligra}{T1}{calligra}{m}{n}
\DeclareFontShape{T1}{calligra}{m}{n}{<->s*[2.0]callig15}{}
\newcommand{\scriptr}{\mathcalligra{r}\,}

\newcommand\inp[2]{\langle #1 \,|\, #2 \rangle}
\newcommand\avg[1]{\langle #1 \rangle}

\DeclareMathAlphabet{\mathpzc}{OT1}{pzc}{m}{it}
	
\definecolor{LightCyan}{rgb}{0.88,1,1}
\definecolor{lightgray}{gray}{0.9}

\def \done      {\textcolor{violet}{Done.}}

\def \intrinsic {\vec{\Lambda}_\mathrm{int}}
\def \extrinsic {\vec{\Lambda}_\mathrm{ext}}
\def \dL        {d_{\rm L}}
\def \Ci        {\mathrm Ci }
\def \Si        {\mathrm Si }

\def \msun      {\rm{M}_\odot}
\def \mchirp    {\mathcal{M}}
\def \imrpd     {\mathtt{IMRPhenomD}}
\def \flow      {f_{\rm low} }

\def \fhigh     {f_{\rm high}}

\def \mchirp    {\mathcal{M}}

\def \pycbc {\textsc{PyCBC}}

\def \bilby {\textsc{BILBY}}
\def \simplepe {\textsc{simple-pe}}
\def \dynesty {\mathtt{dynesty}}

\def \IITGn     {Indian Institute of Technology Gandhinagar, Gujarat 382055, India.\vspace*{5pt}}

\begin{document}

\title{Accelerated parameter estimation of supermassive black hole binaries in LISA \mbox{using a meshfree approximation}}

\author{\textsc{Abhishek~Sharma}\orcidlink{0009-0007-2194-8633}}
\email{sharma.abhishek@iitgn.ac.in }
\affiliation{\IITGn}

\author{\textsc{Anand~S.~Sengupta}\orcidlink{0000-0002-3212-0475}\vspace*{5pt}} 
\email{asengupta@iitgn.ac.in}
\affiliation{\IITGn}

\author{\textsc{Suvodip~Mukherjee}\orcidlink{0000-0002-3373-5236}}
\email{suvodip@tifr.res.in}
\affiliation{Department of Astronomy \& Astrophysics, Tata Institute of Fundamental Research, \\ 1, Homi Bhabha Road, Mumbai- 400005, Maharashtra, India.}

\begin{abstract}
The Laser Interferometer Space Antenna (LISA) will be capable of detecting gravitational waves (GWs) in the milli-Hertz band. Among various sources, LISA will detect the coalescence of supermassive black hole binaries (SMBHBs). Accurate and rapid inference of parameters for such sources will be important for potential electromagnetic follow-up efforts. Rapid Bayesian inference with LISA includes additional complexities as compared to current generation terrestrial detectors in terms of time and frequency dependent antenna response functions. In this work, we extend a recently developed, computationally efficient technique that uses meshfree interpolation methods to accelerate Bayesian reconstruction of compact binaries.  Originally developed for second-generation terrestrial detectors, this technique is now adapted for LISA parameter estimation. Using the full inspiral, merger, and ringdown waveform (PhenomD) and assuming rigid adiabatic antenna response function, we show faithful inference of SMBHB parameters from GW signals embedded in stationary, Gaussian instrumental noise. We discuss the computational cost and performance of the meshfree approximation method in estimating the GW source parameters.
 
\end{abstract}
\maketitle

\section{Introduction}
\label{sec:intro}

The detection of $\sim$100s of gravitational wave (GW) signals by current generation terrestrial detectors, LIGO-Virgo-KAGRA network~\cite{LIGOScientific:2014pky,VIRGO:2014yos,Aso:2013eba}, has firmly established the field of GW astronomy~\cite{KAGRA:2021vkt, Nitz:2021zwj, Mehta:2023zlk}. The current generation of broadband terrestrial interferometric detectors are sensitive in a frequency band that spans a bandwidth from a few Hz to several kHz, and can detect gravitational wave signals from the secular inspiral, merger and ringdown phases of the evolution of compact binary systems composed of neutron stars and stellar-mass black holes out to redshift $\lesssim$1. The Laser Interferometer Space Antenna (LISA), due for its launch in mid 2030s, will open the mHz GW Universe~\cite{LISA:2017pwj}. Many GW sources of interest radiate in this frequency range which include super-massive black hole binaries (SMBHBs) with total mass in the range $10^5-10^8\msun$~\cite{Barausse:2014oca,Klein:2015hvg}, intermediate-mass black hole binaries (IMBHBs) with total mass in the range $10^2-10^5\msun$~\cite{Strokov:2023kmo}, extreme and intermediate mass-ratio inspirals (EMRIs and IMRIs)~\cite{Amaro-Seoane_2007,Babak:2017tow,Gair:2017ynp}, galactic white dwarf binaries (GBs)~\cite{Littenberg:2020bxy}, and stochastic GW background of astrophysical or cosmological origin~\cite{Baghi:2023ast,Babak:2023lro}. Detection and inference of the GW signals from such systems will certainly enhance our understanding of the Universe.

Coalescing MBHBs with a total detector-frame mass between ${10^5 - 10^7\;\msun}$ will be amongst the loudest sources of the GWs in LISA. These sources are expected to be detectable with signal-to-noise ratios (SNRs) of $\mathcal{O}(10^2)-\mathcal{O}(10^3)$ up to high redshifts ($z\sim$$20$)~\cite{LISA:2017pwj,eLISA:2013xep,Barausse:2014oca,Klein:2015hvg}. The formation of a MBHB follows from the merger of two galaxies hosting a MBH at their centers. After the merger of two galaxies, the two MBHs are driven to coalescence from a separation of about a kpc or less. At such large separations, the gravitational wave emission alone will not be sufficient to drive the system to coalescence in Hubble time. For such binaries to form and subsequently merge, there exist three different stages of orbital decay, where a distinct physical mechanism drives the angular momentum and orbital energy loss in each stage~\cite{Begelman:1980vb}. In the first stage, dynamical friction between the MBHs and the medium composed of gas, stars, and dark matter drags the MBHs to the center of the remnant galaxy, where they become gravitationally bound and form a binary. Dynamical friction drags the two MBHs to a separation of $\sim$1$\mathrm{pc}$, subsequently, binary enters the hardening stage where it undergoes three body encounters with the individual stars plunging on nearly radial orbits on the binary~\cite{Bortolas:2021say}. Finally, binary enters the relativistic stage, where the dynamics is dominated by the GW emission. (See references \cite{Dotti:2011um,Mayer:2013jja,Colpi:2014poa,LISA:2022yao} for a detailed discussion of the formation mechanism of MBHBs).

MBHBs can be surrounded by a circumbinary disc, and each BH can be surrounded by separate mini-discs formed out of the gas streaming from the inner edge of the circumbinary disc~\cite{Mayer:2013jja,Colpi:2014poa,DeRosa:2019myq}. As the evolution of the binary takes place in gas rich environments, the accretion of the gas onto BHs can trigger an electromagnetic (EM) counterpart during inspiral, merger and the post-merger phase~\cite{Armitage:2002uu,Milosavljevic:2004cg, Kocsis:2005vv,Dotti:2006zn}. High energy emission originating from the acceration of mini-disc gas onto each MBH can be modulated by the binary orbital motion to an extent that could be observed by future X-ray telescopes~\cite{DalCanton:2019wsr,McGee:2018qwb}. MBHBs detected through GWs serve as standard sirens because they provide direct measurement of the source's luminosity distance. Observation of the associated EM counterpart using telescope facilities is crucial for source's redshift determination. This information can then be used to study the expansion history of the Universe~\cite{Schutz:1986gp,Holz:2005df,DelPozzo:2011vcw, Mukherjee:2019wcg, Balaudo:2022znx} and the propagation properties of GWs over cosmic distances~\cite{PhysRevD.71.084025, Afroz:2024oui}. Moreover, joint GWs and EM observations would impart unique insights on the understanding of accretion physics in violently changing spacetime during the merger.

\begin{figure}[t]
\centering
\includegraphics[width=0.47\textwidth]{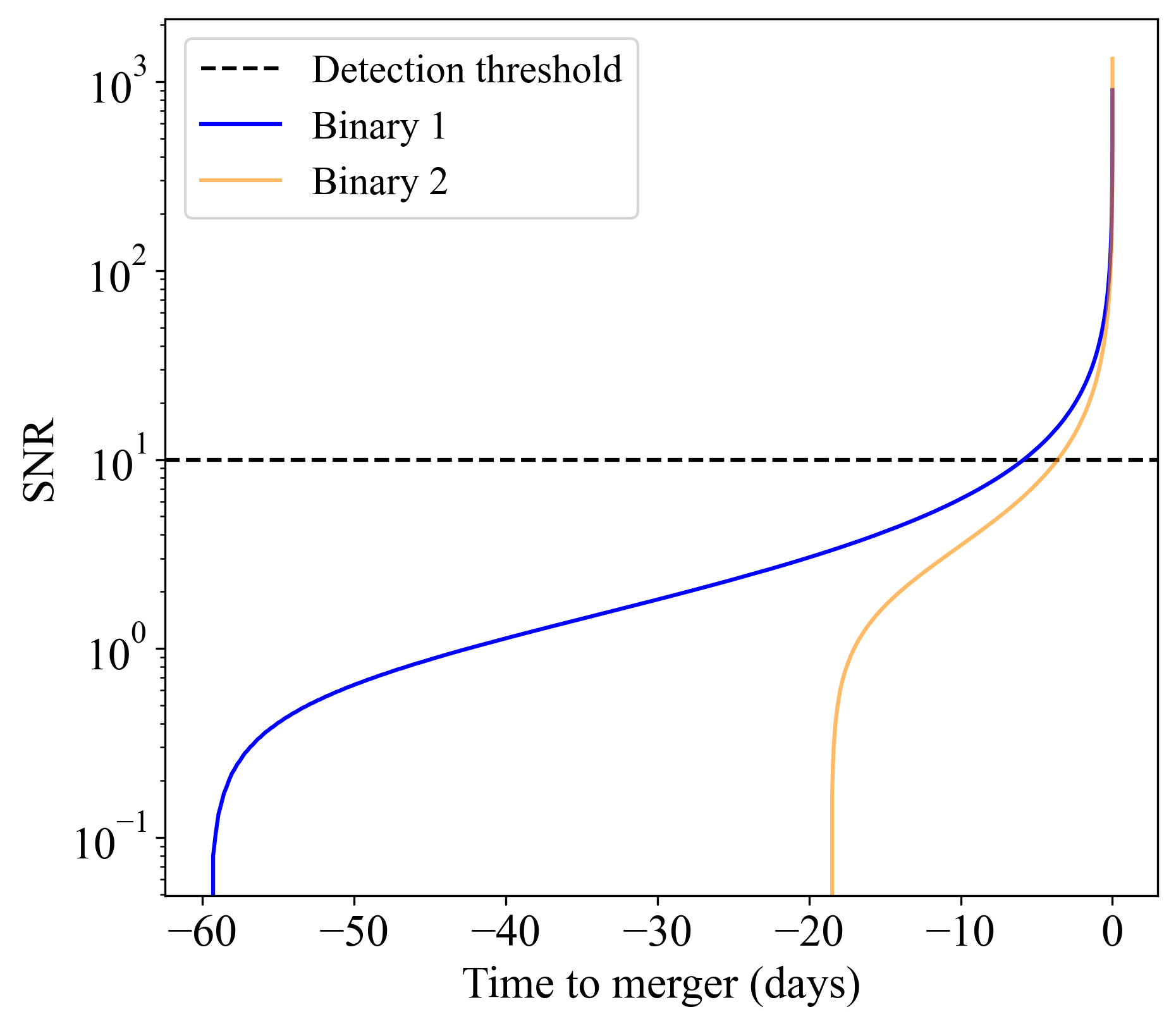}
\caption{Accumulation of the SNR as a function of time for two signals characterized by parameters given in Table \ref{tab:binary_params}. The total SNR for binary 1 and 2 are 903 and 1323, respectively. The horizontal dashed line marks the detection threshold for LISA. Signals from binary 1 and 2 lasts for $\sim$59 and $\sim$19 days in LISA band, but they only become detectable $\sim$6 and $\sim$3.5 days before the merger, respectively. Most of the SNR gets collected just few hours before the merger for both the signals. The noise power spectral densities used to obtain this plot are given in Appendix \ref{sec:noise-psds}.}
\label{fig:snr-vs-t}
\end{figure}

Although GW signals from such systems will last for several weeks in the LISA band, they will only be detectable a few days or hours before the merger, depending on the intrinsic parameters of the system. Fig.\ref{fig:snr-vs-t} shows the accumulation of the SNR as a function of time for MBHBs analyzed in this work. The parameters of the MBHBs are given in Table \ref{tab:binary_params}. We note that the signal lies in the LISA bandwidth for $\sim$59 days and $\sim$19 days for binaries 1 and 2, respectively, but the SNR surpasses the detection threshold $\sim$1 week before the merger and most of the SNR gets collected just few hours before the merger. To facilitate EM follow-up efforts of such sources, improving the computational infrastructure for faster analysis of such signals is crucial. Additionally, rapid parameter inference for such signals enables large-scale inference studies to be conducted within a reasonable amount of time.

Given the data, the source reconstruction problem can be broken into two distinct steps, first being the identification of the merger signal in the data, and second being the subsequent parameter inference of its source. However, the transition to space-based detectors like LISA from the current generation terrestrial detectors brings new challenges in terms of data analysis. Multiple overlapping signals in the LISA data make signal extraction difficult. As signals from MBHBs last for several weeks to months in the LISA band, the detector's motion will need to be considered for unbiased estimation of the source parameters. Additionally, due to expected data gaps, glitches, confusion noise from Galactic binaries, and expected change in the instrument sensitivity over the observation duration, there will be non-stationary and non-Gaussian noise effects. In this work, we only focus on the second step of source reconstruction, i.e., parameter inference of the source. We assume fixed arms of LISA and also assume noise to be Gaussian and stationary.

Historically, the Fisher information matrix (FIM) is considered as an accepted method for determining the uncertainties associated with observable parameters of the source. However, it has been demonstrated that FIM analyses lack the comprehensive information required to make definitive statements about parameter estimation with MBHBs~\cite{Porter:2015eha,Porter:2008kn}. Bayesian inference techniques have been suggested for the searches as well as the parameter estimation of MBHBs in LISA data~\cite{Cornish:2006ms,Cornish:2006ry,Porter:2013wwa,Katz:2024oqg}. A great deal of effort has been put into reducing the computational expense of GW parameter estimation for current generation terrestrial detectors~\cite{Cornish:2010kf,Smith:2013zya,Canizares:2014fya,Vinciguerra:2017ngf,Zackay:2018qdy,Wysocki:2019grj,Morisaki:2020oqk,Qi:2020lfr,Morisaki:2021ngj,Cornish:2021lje,Williams:2021qyt,Hoy:2022tst,Pathak:2022iar,Rose:2022axr,Dax:2022pxd,Lee:2022jpn,Roulet:2022kot,Yelikar:2023mwg,Morras:2023pug,Pathak:2023ixb,Morisaki:2023kuq,Williams:2023ppp,Wong:2023lgb,Tiwari:2023mzf,Fairhurst:2023idl,Tiwari:2024qzr,Wouters:2024oxj}. However, due to the complicated antenna response functions, the estimation of the source parameters for a detector like LISA at low computational cost is a challenging task. Additionally, due to the large expected SNRs of such signals, the likelihood surface exhibits sharp peaks at multiple locations in the degenerate parameter space. Such a kind of likelihood surface poses difficulties for the samplers that utilize MCMC methods to effectively sample the posterior distribution. Recently, several authors have explored the Bayesian analysis of MBHBs with LISA, e.g., the analysis presented in~\cite{Marsat:2020rtl} incorporates fast frequency domain LISA response introduced in~\cite{Marsat:2018oam} and reduced-order model for non-spinning waveforms with higher order modes, the GPU accelerated likelihood approach ~\cite{Katz:2020hku} further extended the analysis incorporating instrument noise. Heterodyned likelihood technique~\cite{Cornish:2010kf,Zackay:2018qdy,Cornish:2021lje} has been widely used for parameter estimation of MBHBs~\cite{Cornish:2020vtw,Cornish:2021smq,Katz:2021uax}. Gravitational wave inference packages like $\pycbc$ and $\bilby$ have also been adapted to incorporate analysis of MBHBs~\cite{Weaving:2023fji,Hoy:2023ndx}. Recently, another rapid parameter estimation algorithm, $\simplepe$~\cite{Fairhurst:2023idl} has been adapted for MBHB parameter estimation~\cite{Hoy:2024ovd}. Several machine learning strategies have also been explored for rapid inference of MBHB parameters~\cite{Du:2023plr,Tang:2024hvk,Vilchez:2024qnw}. In this work, we extend the meshfree approximation method of likelihood interpolation, first introduced in \cite{Pathak:2022iar} for reconstructing compact binaries in the second generation ground-based detector data, for performing analysis with MBHBs in the LISA data. 

The rest of this paper is organized as follows: In Section \ref{sec:method}, we introduce the frequency-averaged antenna pattern functions and the procedure to evaluate the log-likelihood function using meshfree method; in Section \ref{sec:results}, we provide results of our analyses on two MBHBs. We first evaluate the number of frequency bins required to approximate the true antenna pattern functions by performing Bayesian reconstruction of the binary signal injected without noise realization with different number of frequency bins. Later, we perform Bayesian reconstruction of the binaries assuming stationary, Gaussian noise. Finally, we conclude and discuss possible future directions in \mbox{Section \ref{sec:discussion}}. In Appendix \ref{sec:lisa_response}, we provide detailed derivation of the LISA response function used in this work. In Appendix \ref{sec:noise-psds}, we provide the details of power spectral densities for various time-delay interferometry (TDI) channels used in this study.

\section{Meshfree approximation aided Bayesian inference}
\label{sec:method}
\subsection{Likelihood Function and The Frequency Averaged Antenna Pattern}
\label{subsec:Likelihood_function}

The frequency domain response of the LISA detector corresponding to the $k^{\mathrm{th}}$ TDI channel to an incoming GW signal originating from an aligned-spin compact binary system characterized by a set of parameters $\vec{\Lambda} \equiv \{\intrinsic,\; \extrinsic\}$ is given by,
\begin{equation}
    \tilde{h}^{(k)}(f; \vec{\Lambda}) = C_{(k)}(f; \extrinsic)\:e^{-2 \pi i f \Delta t}\:\tilde{h}_+(f; \intrinsic),
    \label{eq:Waveform-in-TDI1}
\end{equation}
where, $k \in [A, E, T]$; $\intrinsic$ denotes the parameters intrinsic to the source such as component masses and spins, while $\extrinsic$ denotes the extrinsic parameters, namely, ecliptic longitude ($\lambda$), ecliptic latitude ($\beta$), polarization angle ($\psi$), inclination angle ($\iota$) and the luminosity distance ($\dL$). $\tilde{h}_+(f; \intrinsic)$ denotes the Fourier transform of the GW signal as measured at the solar system barycenter (SSB) (this term also includes TDI-induced pre-factor, see Eq.\eqref{eq:h_tdi}) and $\Delta t$ corresponds to the extra time taken by GW signal to travel from SSB to the centroid of LISA triangle. And $C_{(k)}(f; \extrinsic)$ is the complex valued frequency-series given by,   
\begin{equation}
    C_{(k)}(f; \extrinsic) = \frac{1}{\dL}\big[F_{(k)}^+(f; \vec{\alpha}) A_+(\iota) + i F_{(k)}^\times(f; \vec{\alpha}) A_\times(\iota)\big],
\end{equation}
where, $\vec{\alpha} = (\lambda, \beta, \psi)$; $F_{(k)}^{+,\times}(f; \vec{\alpha})$ are the antenna pattern functions corresponding to the plus and cross polarizations of the GW signal for $k^{\mathrm{th}}$ TDI channel.
$A_+$ and $A_\times$ are the inclination dependent terms (see Appendix \ref{sec:lisa_response} for details).

Given the data $\mathbf{d}(t) = \mathbf{s}(t, \vec{\Lambda}_{\mathrm{true}}) + \mathbf{n}(t)$ containing a possible GW signal $\mathbf{s}(t)$ characterized by the parameters $\vec{\Lambda}_{\mathrm{true}}$ in additive stationary and Gaussian noise $\mathbf{n}(t)$, we are interested in solving the stochastic inverse problem by estimating the posterior distribution $p(\vec{\Lambda}\mid \mathbf{d})$ of the source parameters. Bayes' theorem relates the posterior distribution to the likelihood function $\mathcal{L}(\mathbf{d}\mid\vec{\Lambda})$ of observing the data $\mathbf{d}$ given the source parameters are $\vec{\Lambda}$ as follows,
\begin{equation}
    p(\vec{\Lambda}\mid \mathbf{d}) = \frac{\mathcal{L}(\mathbf{d}\mid\vec{\Lambda})p(\vec{\Lambda})}{p(\mathbf{d})},
    \label{eq:Bayestheorem}
\end{equation}
where, $p(\vec{\Lambda})$ is the prior probability distribution over the parameters and $p(\mathbf{d})$ is the Evidence which act as an overall normalization constant. Our parameter space is composed of eight parameters which include two intrinsic parameters, namely, the component masses $(m_{1,2})$, and six extrinsic parameters namely, ecliptic longitude $(\lambda)$, ecliptic latitude $(\beta)$, polarization angle $(\psi)$, inclination $(\iota)$, luminosity distance $(d_\mathrm{L})$ and time of coalescence $(t_c)$. We consider the binary to be spinless and fix spin magnitudes to zero while generating our simulated signals. Assuming noise to be Gaussian and stationary, the phase marginalized log-likelihood function can be written as follows~\cite{Thrane:2018qnx}:
\begin{align}
    \nonumber
    \ln \mathcal{L}\left(\boldsymbol{d}^{(k)} \mid \vec{\Lambda}\right)= &\ln I_0\left[\left|\sum_{k=A,E,T}\left\langle\boldsymbol{d}^{(k)} \mid h^{(k)}(\vec{\Lambda})\right\rangle\right|\right] \\
    & -\frac{1}{2} \sum_{k=A,E,T}\left\langle h^{(k)}(\vec{\Lambda}) \mid h^{(k)}(\vec{\Lambda})\right\rangle,
    \label{eq:logl}
\end{align}
where, $I_0[\cdot]$ is the zeroth-order modified Bessel function of the first kind, index $k$ denotes the TDI channels ${A, E, T}$ and $\inp{a}{b}$ is the noise-weighted inner product of two time domain signals $a(t)$ and $b(t)$ in $k^{\mathrm{th}}$ channel,
\begin{equation}
    \displaystyle \inp{a}{b} = 4 \:\Re\int_{\flow}^{\fhigh} \frac{\tilde{a}^{(k)\ast}(f) \: \tilde{b}^{(k)}(f)}{S_n^{(k)}(f)}\: df,
    \label{eq:innerProduct}
\end{equation}
where $S_n^{(k)}(f)$ is the one-sided noise power spectral density of the $k^{\mathrm{th}}$ TDI channel. The PSD for various TDI channels used in this work are obtained from LDC manual~\cite{Sangria_LDC} and are given in Appendix \ref{sec:noise-psds}.

For fast computation of the log-likelihood function a meshfree interpolation is carried out only over the intrinsic parameter space $(\intrinsic)$. For this purpose, it is imperative to factorize the signal response in the detector given by Eq.\eqref{eq:Waveform-in-TDI1} into two pieces --- one that only depends on the intrinsic parameters of the signal and an overall amplitude factor that depends only on the extrinsic parameters. For aligned-spin waveform models, such is the case for the signal response in current generation terrestrial detectors like advanced LIGO, where the long-wavelength approximation is valid and the signal duration is very short as compared to the rotational timescale of the earth, so that the antenna pattern can be taken as a constant factor. Under such a scheme, the overlap integrals appearing on the RHS of Eq.\eqref{eq:logl} only have to be evaluated over $\intrinsic$; while the extrinsic parameter dependent part is cheap to calculate and gets multiplied as an overall amplitude factor. However, for LISA, given the form of $C_{(k)}(f; \extrinsic)$, decoupling extrinsic parameters from the integral poses a challenge as $C_{(k)}$ explicitly depends on the frequency. To overcome this challenge, we approximate the signal response by dividing the whole frequency bandwidth into $n$ sub-bands and taking the analytic average of $C_{(k)}(f; \extrinsic)$ over frequency in each sub-band. The signal response in a particular sub-band (let's say $j^{\mathrm{th}}$ sub-band) can be written as,
 \begin{equation}
    \label{eq:Waveform-in-TDI2}
    \tilde{h}^{(k)}(f; \vec{\Lambda}) \approx \avg{C_{(k)}(\extrinsic)}_{(f_j, f_{j+1})} \:e^{-2 \pi i f \Delta t}\tilde{h}_+(f; \intrinsic),
 \end{equation}
 where, $f\in [f_j, f_{j+1}]$ and $\avg{C_{(k)}(\extrinsic)}_{(f_j, f_{j+1})}$ denotes a complex constant approximating the antenna response function in the $j^{\mathrm{th}}$ sub-band. Thus, the antenna response across the entire bandwidth can be approximated as a piece-wise constant function. The integral in Eq.\eqref{eq:innerProduct} over the full bandwidth $[f_{\mathrm{low}}, \;f_{\mathrm{high}}]$ can be broken into $n$ separate integrals calculated over each sub-band, i.e.,
 \begin{equation}
     \label{eq:integral_in_bands}
     \int_{\flow}^{\fhigh} \frac{\tilde{a}^{(k)\ast}(f) \: \tilde{b}^{(k)}(f)}{S_n^{(k)}(f)}\: df = \sum_{j=0}^{n-1} \int_{f_j}^{f_{j+1}} \frac{\tilde{a}^{(k)\ast}(f) \: \tilde{b}^{(k)}(f)}{S_n^{(k)}(f)}\: df,
 \end{equation}
 where, $f_0 = f_{\mathrm{low}}$ and $f_n = f_{\mathrm{high}}$.
 
\begin{figure}[t]
\centering
\includegraphics[width=0.48\textwidth]{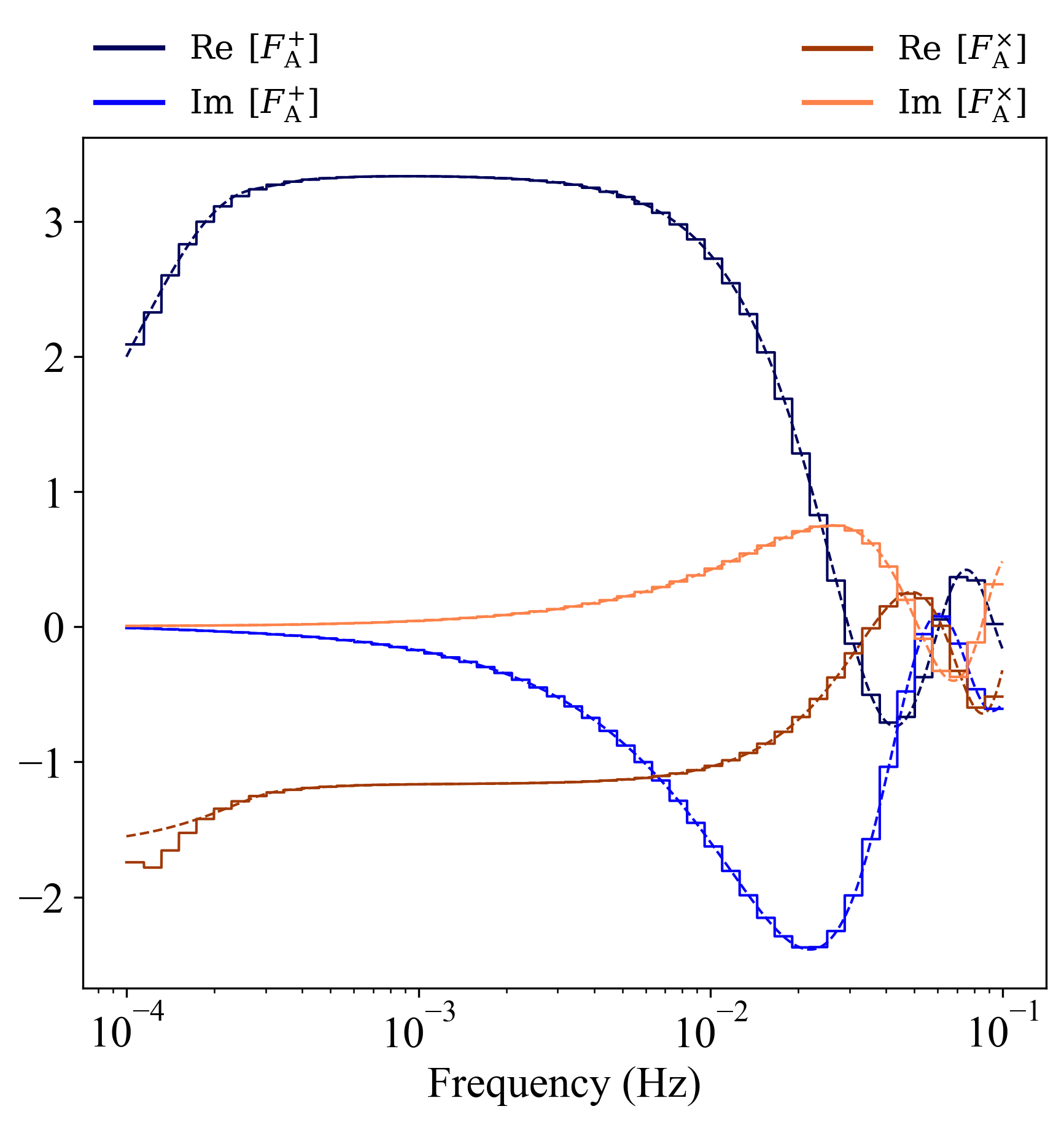}
\caption{Comparison of the real and imaginary parts of the approximated and actual antenna pattern functions for the two polarizations for TDI channel A. The approximated antenna pattern functions are calculated by averaging transfer functions (Eq.\eqref{eq:average-general}) over several frequency intervals. A total of 50 frequency intervals distributed uniformly in the logarithmic scale are created and averaged transfer functions are evaluated in each bin. The dashed curves shows the actual antenna pattern functions without any approximation and the solid steps shows the antenna pattern functions averaged over the frequency.}
\label{fig:antenna-pattern}
\end{figure}
The quantities that explitly depend on the frequency in the expressions of antenna response functions are the transfer functions, $\mathcal{T}_{sr}(f, \lambda, \beta)$ (see Eqs.\eqref{eq:transfer_function} and \eqref{eq:G_X}). Therefore, averaging $C_{(k)}(f; \extrinsic)$ with respect to frequency results in averaging the transfer functions $\mathcal{T}_{sr}(f, \lambda, \beta)$. The average over frequency of $\mathcal{T}_{sr}(f)$ in the interval $[f_{\mathrm{l}}, \;f_{\mathrm{h}}]$ can be evaluated by writing their real and imaginary parts separately and thereafter, calculating average of the individual components using average value theorem, i.e.,
\begin{align}
    \label{eq:average-general}
    \nonumber
     \avg{\mathcal{T}_{sr}(\lambda, \beta)}_{(f_l, f_h)} &= \avg{\Re[\mathcal{T}_{sr}(\lambda, \beta)]}_{(f_l, f_h)} \\
     \nonumber
     &+ i \avg{\Im[\mathcal{T}_{sr}(\lambda, \beta)]}_{(f_l, f_h)} \\
     \nonumber
     \avg{\mathcal{T}_{sr}(\lambda, \beta)}_{(f_l, f_h)} &= \frac{1}{f_h-f_l}\bigg[\int_{f_l}^{f_h}\Re[\mathcal{T}_{sr}(f, \lambda, \beta)] \: df \\
    & + i \int_{f_l}^{f_h}\Im[\mathcal{T}_{sr}(f, \lambda, \beta)] \: df\bigg].
\end{align}
The integrals appearing in the above equation are given by the following closed-form expressions:
\begin{widetext}
    \begin{align}
        \label{eq:avg_Re_T}
        \int \Re[\mathcal{T}_{sr}(f, \lambda, \beta)] df &= -\frac{f_\star}{k_{sr}^2 - 1}\bigg[(k_{sr} - 1)\Si\big(\frac{f k_s}{\sqrt{3}f_\star}\big) + (1 + k_{sr}) \Si\big(\frac{f}{3f_\star}(6 + \sqrt{3}k_s)\big) - 2k_{sr} \Si\big(\frac{f}{3f_\star} (3 + 3k_{sr} + \sqrt{3}k_s)\big)\bigg], \\
        \label{eq:avg_Im_T}
        \int \Im[\mathcal{T}_{sr}(f, \lambda, \beta)] df &= -\frac{f_\star}{k_{sr}^2 - 1}\bigg[(k_{sr} - 1)\Ci\big(\frac{f k_s}{\sqrt{3}f_\star}\big) + (1 + k_{sr}) \Ci\big(\frac{f}{3f_\star}(6 + \sqrt{3}k_s)\big) - 2k_{sr} \Ci\big(\frac{f}{3f_\star} (3 + 3k_{sr} + \sqrt{3}k_s)\big)\bigg],
    \end{align}
\end{widetext}
where, $k_{sr} = \hat{k}\cdot\hat{\scriptr}_{sr}$, $k_s = \hat{k}\cdot\hat{\scriptr}_{s}$; $\hat{k}$ denotes the direction of the GW propagation, ${\hat{\scriptr}_{sr}}$ is a unit vector pointing from sending spacecraft $s$ to the receiving spacecraft $r$, ${\hat{\scriptr}_{s}}$ is a unit vector pointing in the direction of spacecraft $s$ from the centroid of LISA, and $\Si(x)$ and $\Ci(x)$ are the Sine and Cosine integral functions respectively and are given by,

\begin{align}
    \label{eq:Sine_integral}
    &\Si(x) = \int_0^x \frac{\sin(t)}{t} dt \\
    \label{eq:Cosine_integral}
    \Ci(x) = \gamma &+ \log(x) + \int_0^x \frac{\cos(t) - 1}{t} dt,
\end{align}
where, $\gamma$ is the Euler-Mascheroni constant. Thus, the log-likelihood function in Eq.~\eqref{eq:logl} can be written as,
%
%
\begin{widetext}
\begin{multline}
    \label{eq:logl2}
    \ln \mathcal{L}\left(\boldsymbol{d}^{(k)} \mid \vec{\Lambda}\right) = \ln I_0\left[\left|\sum_{k=A,E,T}\sum_{j=0}^{n-1}\avg{C_{(k)}(\extrinsic)}_{(f_j, f_{j+1})}\left\langle\boldsymbol{d}^{(k)} \mid h_+(\intrinsic)\right\rangle_{(f_j, f_{j+1})}\right|\right] \\
    -\frac{1}{2} \sum_{k=A,E,T}\sum_{j=0}^{n-1}\norm{\avg{C_{(k)}(\extrinsic)}_{(f_j, f_{j+1})}}^2\left\langle h_+(\intrinsic) \mid h_+(\intrinsic)\right\rangle_{(f_j, f_{j+1})}.
\end{multline}
\end{widetext}

In Fig~\ref{fig:antenna-pattern} we show the comparison of the frequency dependant antenna pattern functions, $F_A^+$ and $F_A^\times$ calculated without any approximation with the piece-wise constant approximation as described above. The figure shows 50 frequency bins uniformly spaced on the logarithmic scale within the entire bandwidth. For each frequency bin the averaged transfer functions are calculated using Eqs.~\eqref{eq:average-general}, \eqref{eq:avg_Re_T} and \eqref{eq:avg_Im_T} and finally the antenna pattern functions are evaluated using Eqs.~\eqref{eq:G_X} and \eqref{eq:AETchannels}. The bins are placed uniformly on a logarithmic scale due to the characteristic time-frequency track of the binary. As in the response function the position of spacecrafts are calculated at the time corresponding to a given value of frequency as obtained from Eq.~\eqref{eq:t-fmap}. Logarithmic-uniform spaced bins in frequency make sure relatively uniform spacing in time and thus the approximated antenna response function mimics the original one. Note that the approximated antenna pattern slightly deviates from the true one in the smaller frequency region. However, due to the insignificant contribution of the SNR in that frequency range, this deviation can be ignored.

\subsection{Meshfree Likelihood Interpolation}
\label{subsec:Meshfree_likelihood_interpolation}
Meshfree interpolation of the likelihood function was first introduced in \cite{Pathak:2022iar} for reconstructing compact binaries in a single second generation ground-based detector data. The analysis was further extented to a network of detectors in \cite{Pathak:2023ixb}. The meshfree method of likelihood interpolation comprises two distinct stages:
\begin{enumerate}
    \item Start-up stage: During this stage, the radial basis function (RBF) interpolants are created for the relevant quantities.
    \item Online stage: The likelihood function is evaluated at arbitrary query point proposed by the sampling algorithm using the RBF interpolants.
\end{enumerate}
\subsubsection{Start-up stage}
\label{subsubsec:start-up}
In this stage, we generate RBF interpolants for the relevant quantities, which are used to quickly evaluate the likelihood function during the online stage. We first place N number of nodes in the intrinsic parameter space $\{\intrinsic\}$, denoted by $\intrinsic^n: \:\:n\in(1, N)$.  The intrinsic parameter space in our case is two-dimensional and we take $N=800$. Nodes are uniformly distributed within a moderate-sized rectangular boundary in $(\mchirp, q)$ coordinates centered around the best matched template point known from an upstream search. In this work, we take the injected intrinsic parameters as the best matched template point. The choice of nodes and the boundary in the intrinsic parameter space where nodes are to be placed impact the quality of the derived basis vectors~(see Eq. \eqref{eq:SVD}). If the nodes are placed very far away from the peak of the likelihood function they do not contribute much to improve the interpolation accuracy. A metric guided placement of random nodes can be shown to improve accuracy with fewer nodes.

We evaluate the overlap integrals, namely,
\begin{align}
    \nonumber
    \vec{z}^{(k)}_j(\intrinsic, t_c) &= \left\langle\boldsymbol{d}^{(k)} \mid h_+(\intrinsic)\right\rangle(t_c)_{(f_j, f_{j+1})} \\
    &= \int_{f_j}^{f_{j+1}} \frac{\boldsymbol{\tilde{d}}^{(k)*}(f) \: \tilde{h}_+(\intrinsic, f)}{S_n^{(k)}(f)} e^{-2 \pi i f t_c}\: df,
    \label{eq:z_vec}
\end{align}
and,
\begin{align}
    \nonumber
    \sigma^2(\intrinsic)^{(k)}_j 
    &= \left\langle h_+(\intrinsic, t_c) \mid h_+(\intrinsic, t_c)\right\rangle_{(f_j, f_{j+1})} \\
    &= \int_{f_j}^{f_{j+1}} \frac{ \left | \tilde{h}_+(\intrinsic, f) \right |^2}{S_n^{(k)}(f)} \: df,
    \label{eq:sigma_square}
\end{align}
at each of the RBF nodes $\intrinsic^n$ for every frequency bin. Note that the index $k$ denotes the TDI channel A, E or T and index $j$ denotes the frequency bin. The overlap integral between the data $\boldsymbol{d}^{(k)}$ and $h_+(\intrinsic)$ in Eq.~\eqref{eq:z_vec} is evaluated at discrete values of the circular time shifts $t_c$ uniformly spaced over a specified range centered around the trigger time and thus represented as a vector quantity. The range of $t_c$ spans $\pm 30$ minutes around the trigger time. Note that the template waveform, $h_+(\intrinsic)$ used to calculate the time-series, $\vec{z}^{(k)}_j(\intrinsic^n)$ are given (in frequency domain) by Eq.\eqref{eq:h_tdi}. We would like to identify a set of suitable basis vectors that span the space of $N\:$ $\vec{z}^{(k)}_j(\intrinsic^n)$ for each $j$. For a single frequency bin, this is performed by stacking these vectors horizontally to form a matrix $\mathcal{Z}^{(k)}_j$, where each row corresponds to a single node. Subsequently, performing the Singular Value Decomposition (SVD) of the resultant matrix yields the desired basis vectors $\vec{u}_{\mu, j}^{(k)}$, where $\mu \in (1, N)$. Any row of the matrix $\mathcal{Z}^{(k)}_j$ can be expressed as a linear combination of these basis vectors $\vec{u}_{\mu, j}^{(k)}$,
\begin{equation}
    \vec{z}^{(k)}_j(\intrinsic^n) = \sum_{\mu=1}^N \; C_{\mu,j}^{n \;(k)}\; \vec{u}_{\mu,j}^{(k)},
    \label{eq:SVD}
\end{equation}
where, $C_{\mu,j}^{n \;(k)} \equiv C_{\mu,j}^{(k)}(\intrinsic^n)$, $\mu \in (1, N)$ are $N$ SVD coefficients. SVD makes sure that the set of orthonormal basis vectors $\vec{u}_{\mu,j}^{(k)}$ are arranged in decreasing order of their relative importance as determined by the spectrum of singular values. Only first few $\ell$ basis vectors contribute to the summation in Eq.~\eqref{eq:SVD} $(\ell \ll N)$. The coefficients $C_{\mu,j}^{n \;(k)}$ are smooth functions over $\intrinsic^n$ and therefore, at any point in the parameter space $\intrinsic^q$ can be expressed as a linear combination of the radial basis functions~\cite{doi:10.1142/6437} (RBFs) centered at the interpolation nodes,
\begin{equation}
    C_{\mu,j}^{q(k)}=\sum_{\gamma=1}^N a_{\gamma, j}^{(k)} \phi_{j}\left(\left\|\intrinsic^q-\intrinsic^\gamma\right\|_2\right)+\sum_{m=1}^M b_{m, j}^{(k)} p_{m, j}\left(\intrinsic^q\right),
    \label{eq:C_interpolants}
\end{equation}
where, $\phi_{j}$ is the RBF kernel centered at $\intrinsic^n \in \mathbb{R}^2$, and $\{p_{m, j}\}$ denotes the monomials that span the space of polynomials of some predefined degree $\nu$ in 2 variables. $\sigma^2(\intrinsic^q)^{(k)}_j$ is also a smooth scalar field over $\intrinsic^n$ thus, can be similarly expressed in terms of the RBF functions and the monomials. In order to evaluate $C_{\mu,j}^{q(k)}$ at an arbitrary query point, Eq.~\eqref{eq:C_interpolants} implies that we need a total of $(N+M)$ coefficients. The SVD coefficients $C_{\mu,j}^{q \;(k)}$ and $\sigma^2(\intrinsic^q)^{(k)}_j$ are known at the N interpolation nodes, thus provide N interpolation conditions. In order to uniquely determine the coefficients $a_{\gamma, j}$ and $b_{m, j}$, we impose M extra conditions $\sum_{m=1}^M a_{m, j}^{(k)}p_{m, j}(\intrinsic^q) = 0$. This results in the following system of linear equations for each frequency bin,
\begin{equation}
\left[\begin{array}{cc}
\boldsymbol{\Phi} & \boldsymbol{P} \\
\boldsymbol{P}^T & \boldsymbol{O}
\end{array}\right]\left[\begin{array}{l}
\boldsymbol{a}^{(k)} \\
\boldsymbol{b}^{(k)}
\end{array}\right]=\left[\begin{array}{c}
C_\mu^{n(k)} \\
\mathbf{0}
\end{array}\right],
\label{eq:constraints}
\end{equation}
where, the components of the matrices $\boldsymbol{\Phi}$ and $\boldsymbol{P}$ are ${\Phi_{ab} = \phi\left(\left\|\intrinsic^a-\intrinsic^b\right\|_2\right)}$ and $P_{ab} = p_b(\intrinsic^a)$, respectively; $\boldsymbol{O}_{M\times M}$ is a zero-matrix and $\mathbf{0}_{M\times 1}$ is a zero-vector. The constraint equations given by Eq.~\eqref{eq:constraints} can be uniquely solved for the $(N+M)$ unknown coefficients $\boldsymbol{a}^{(k)}$ and $\boldsymbol{b}^{(k)}$, thus determining the meshfree interpolants in Eq.~\eqref{eq:C_interpolants}. As mentioned above, due to the rapidly decreasing singular values, only first few $\ell$ basis vectors are required in the summation Eq.~\eqref{eq:SVD}. Thus, in order to evaluate $\vec{z}^{(k)}_j$ at any point in the intrinsic parameter space we need to generate $\ell$ interpolants and a single interpolant is generated for $\sigma^{2(k)}_j$. Therefore, a total of $(\ell + 1)$ interpolants are required to be generated for each frequency bin. Once these interpolants are generated, we move on to the online stage where likelihood can be computed at an arbitrary point proposed by the sampler.
\subsubsection{Online stage}
\label{subsubsec:Online}
Using the meshfree interpolants created in the start-up stage, we can evaluate the coefficients $C_{\mu,j}^{q(k)}$ and $\sigma^2(\intrinsic^q)^{(k)}_j$ for $j^{\mathrm{th}}$ frequency bin at an arbitrary query point $\intrinsic^q$ promptly. Corresponding $\vec{z}^{(k)}_j(\intrinsic^q)$ can be evaluated using Eq.~\eqref{eq:SVD} restricting the summation upto first $\ell$ terms. However, because of the finite sample rate, it may happen that none of the components of $\vec{z}^{(k)}_j(\intrinsic^q)$ may correspond to the query time $t_c^\ell$, the trigger time as observed in the LISA frame. Therefore, we take 10 time samples centered around the sample nearest to $t_c^\ell$ and fit these samples with a cubic-spline. Subsequently, we evaluate the $z^{(k)}_j(\intrinsic^q)$ at $t_c^\ell$ using the cubic-spline interpolant. Once the interpolated values of the quantities $z^{(k)}_j$ and $\sigma^{2(k)}_j$ are evaluated for all the values of $j$ at $\intrinsic^q$, the log-likelihood can be evaluated using Eq.\eqref{eq:logl2}.

While generating the interpolants for the coefficients $C_{\mu,j}^{n \;(k)}$ in the intrinsic parameter space, the width of the boundary chosen along chirp mass and mass ratio affect the faithfulness of the interpolation. Arbitrarily large boundaries result in highly oscillatory coefficient surfaces, thus lead to the generation of unfaithful interpolants for a given number of nodes. Therefore, the size of the rectangular boundary in $(\mchirp, q)$ plane is taken such that reliable interpolants could be generated. We restrict the width $\Delta\mchirp$ and $\Delta q$ of the rectangular boundary such that corresponding widths in dimensionless chirp time coordinates $(\Delta\theta_0, \Delta\theta_3) \sim (\pm2.5, \pm2.5)$ around the central value (see \cite{Owen:1998dk} for the definition of dimensionless chirp time coordinates).
\section{Results}
\label{sec:results}

\begin{table*}[t]
    \centering
    \begin{tabular}{c c c c c c c c c c c}
    \toprule[1pt]
        Binary \ \ & $m_1 \: (10^6 \msun)$ \ \ & $m_2 \: (10^6 \msun)$ \ \ & $d_L \: (\mathrm{Gpc})$ \ \ & $\iota $ \ \ & \ \ \ $\lambda $ \ \ \ & \ \ \ $\beta $ \ \ \ & \ \ \ $\psi $ \ \ \ & \ \ \ $t_c \: (\mathrm{s})$ \ \ \ & \ \ \ $\rho_{\mathrm{opt}}$ \ \ \ & CPU hours \\
        
        \midrule[1pt]
        1  & 1.5 & 0.5 & 36.68 & 1.05 & 0.94 & 1.02 & 2.03 & 18144000 & 903 & $\sim$900\\

        2  & 3.0 & 1.0 & 36.68 & 1.05 & 0.94 & 1.02 & 2.03 & 18144000 & 1323 & $\sim$858 \\
    
    \bottomrule[1pt]
    \bottomrule[1pt]
    \end{tabular}
    \caption{Injection parameters of each binary. The second last column shows the optimal SNR of the GW signal. The last column shows the computational time taken by each PE run in obtaining the posterior samples. Note that the component masses are in the detector frame. The extrinsic parameters, namely the coalescence time, ecliptic longitude, ecliptic latitude and polarization angle are given in the SSB frame. The injected value of luminosity distance corresponds to the redshift of 4 assuming .}
    \label{tab:binary_params}
\end{table*}

\begin{table}[t]
    \centering
    \begin{tabular}{c l c}
    \toprule[1pt]
        Parameters \ \ & Prior distribution \ \ & \ \ \ Range \\
        
        \midrule[1pt]
        $\mchirp$  & $\propto \mchirp$ & $[\mchirp_{\mathrm{cent}} \pm 500 \msun]$ (Binary 1)\\

          &  & $[\mchirp_{\mathrm{cent}} \pm 2000 \msun]$ (Binary 2)\\
        
        $q$  & $\propto [(1 + q)/q^3]^{2/5}$ & $[q_{\mathrm{cent}} \pm 0.2]$ \\

        $\dL$  & Uniform & $[3, 60] (\mathrm{Gpc})$ \\

        $t_c^\ell$  & Uniform & $t_{\mathrm{trig}} \pm 1800s$ \\

        $\lambda_\ell$  & Uniform & $[0, 2\pi]$ \\

        $\beta_\ell$  & Uniform in $\sin(\beta)$ & $[-\pi/2, \pi/2]$ \\

        $\iota$  & Uniform in $\cos(\iota)$ & $[0, \pi]$ \\

        $\psi_\ell$  & Uniform & $[0, \pi/2]$ \\
    \bottomrule[1pt]
    \bottomrule[1pt]
    \end{tabular}
    \caption{Priors used in Bayesian inference over various parameters. The chosen prior distributions over $\mchirp$ and $q$ corresponds to uniform distributions over component masses. Priors over ecliptic longitude and latitude corresponds to isotropic direction across all sky. The superscript (subscript) `$\ell$' on $t_c$ ($\lambda$, $\beta$ and $\psi$) denotes that these parameters are measured in the LISA constellation frame.}
    \label{tab:priors}
\end{table}

\begin{figure*}[t]
\centering
\includegraphics[width=\textwidth]{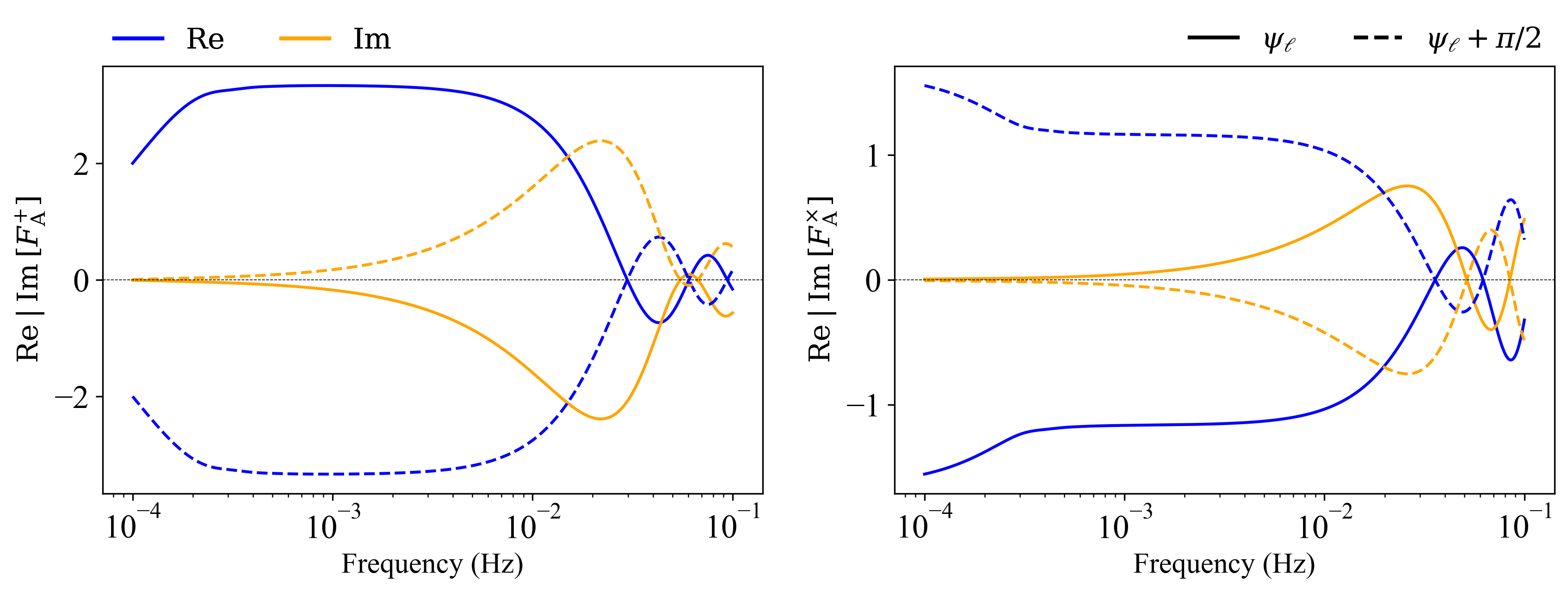}
\caption{The real and imaginary parts of the TDI-A antenna pattern functions corresponding to plus (left plot) and cross (right plot) polarization states for binary 1. The polarization angle for binary 1 in LISA frame, $\psi_\ell = 1.25$. The blue and orange curves represent real and imaginary parts of the antenna patterns, respectively. The solid lines corresponds to $\psi_\ell$ and the dashed lines corresponds to $\psi_\ell + \pi/2 = 2.82$. Note that the antenna patterns flip sign under transformation $\psi_\ell\rightarrow\psi_\ell + \frac{\pi}{2}$.}
\label{fig:polarization-degeneracy}
\end{figure*}

To test the meshfree method of likelihood interpolation for reconstructing the MBHB source parameters, we consider two sources with parameters given in Table \ref{tab:binary_params}. We fabricate data streams in TDI channels A, E and T sampled at $0.2\: \mathrm{Hz}$ for these two sources assuming stationary, Gaussian noise coloured with the TDI channel specific noise power spectral density (PSD). We use the $\imrpd$ waveform approximant~\cite{PhysRevD.93.044007} to model the GW signal from MBHB. The frequency bandwidth considered in this analysis range from $10^{-4}\;\mathrm{Hz}$ to $10^{-1}\;\mathrm{Hz}$. We sample the posterior distribution in chirp mass $(\mchirp)$, mass ratio $(q)$, luminosity distance $(\dL)$, time at coalescence $(t_c^\ell)$, ecliptic longitude $(\lambda_\ell)$, ecliptic latitude $(\beta_\ell)$, inclination $(\iota)$, and polarization angle $(\psi_\ell)$. The priors used for various parameters in our Bayesian analysis are given in Table \ref{tab:priors}. The priors for chirp mass and mass ratio are such that the transformed joint distribution over component masses is uniform. As labelling MBHB signals in terms of time at coalescence measured in the SSB frame is a bad parametrization~\cite{Marsat:2020rtl}. The extrinsic parameters namely the time at coalescence, ecliptic longitude, ecliptic latitude and polarization angle are sampled in the LISA constellation frame instead of the SSB frame. The conversion of these parameters from the SSB frame to the LISA frame is given in ~\cite{Marsat:2020rtl}. While sampling in the LISA constellation frame there exist eight degeneracies in the extrinsic parameter space, four longitudinal at $\left\{\lambda_\ell+\frac{\pi}{2}(0,1,2,3), \psi_\ell+\frac{\pi}{2}(0,1,2,3)\right\}$ and two latitudinal at $\{ \pm \beta_\ell, \pm \cos \iota, \pm \cos \psi_\ell\}$. However, due to the time and frequency dependence of the antenna pattern functions, considered in our analysis, these degeneracies may get broken. These degeneracies are discussed in great detail in \cite{Marsat:2020rtl}. Furthermore, under the transformation $\psi_\ell\rightarrow\psi_\ell + \frac{\pi}{2}$, the antenna pattern functions $F_\mathrm{A, E, T}^{+, \times}$ acquire an overall minus sign as shown in Fig. \ref{fig:polarization-degeneracy}. The log-likelihood function given by Eq.~\eqref{eq:logl} is degenerate with respect to this transformation. Therefore, utilizing the degeneracy with respect to polarization angle, we take the interval $[0, \frac{\pi}{2}]$ as prior width for the polarization angle, $\psi_\ell$. This helps in reducing the parameter space volume to be explored by the sampler and thus facilitates faster convergence towards the maximum likelihood regions. The restricted prior range in $\psi_\ell$ is sufficient to recover parameters of the binaries having $\psi_\ell \in [0, \pi]$. The prior volume in the sky location parameters namely, ecliptic longitude $(\lambda_\ell)$ and ecliptic latitude $(\beta_\ell)$ span the whole sky. 

\begin{figure*}[t]
\centering
\includegraphics[width=\textwidth]{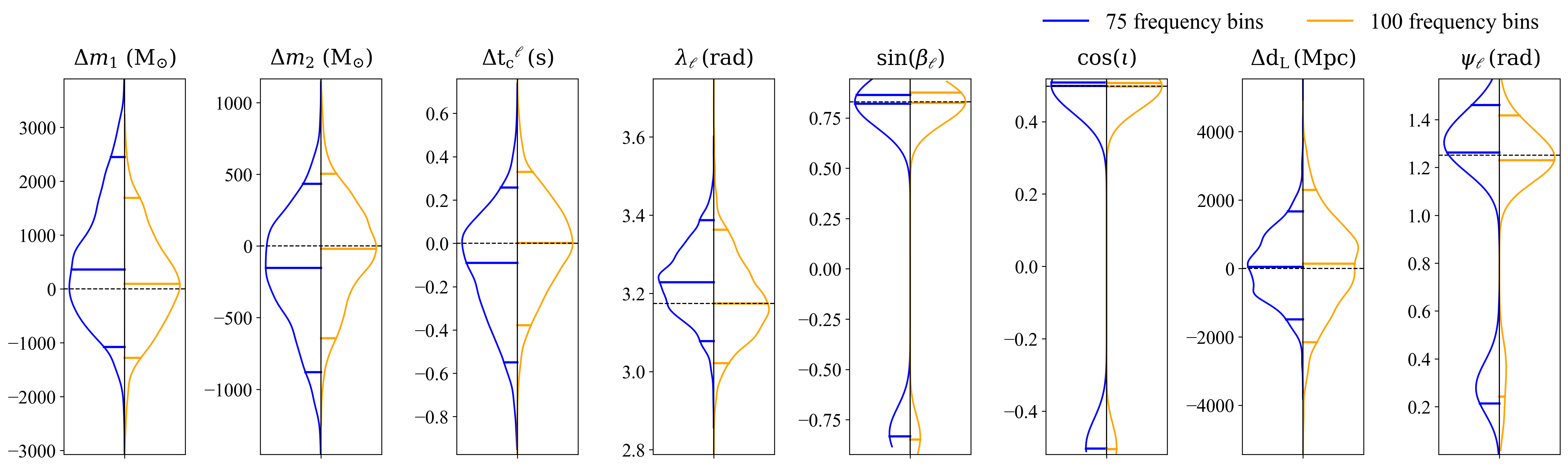}
\caption{Comparison of the marginalized posterior distributions for MBHB 1 of Table \ref{tab:binary_params}, injected without noise realization obtained using 75 and 100 frequency bins. The horizontal dashed black line mark the true value of the parameter. All the parameters are recovered within their 90\% CI in both cases, but inference is better with 100 frequency bins. Median values of the parameters lie closer to the injected values and the posterior width for the component masses is also narrower with 100 frequency bins. Note that in this, and the following plots the extrinsic parameters namely, ecliptic longitude, ecliptic latitude, polarization angle, and the time at coalescence are shown in LISA frame, and the difference with respect to the injected values are plotted for component masses, time at coalescence, and luminosity distance.}
\label{fig:binary1-75vs100bins}
\end{figure*}

\begin{figure*}[t]
\centering
\includegraphics[width=\textwidth]{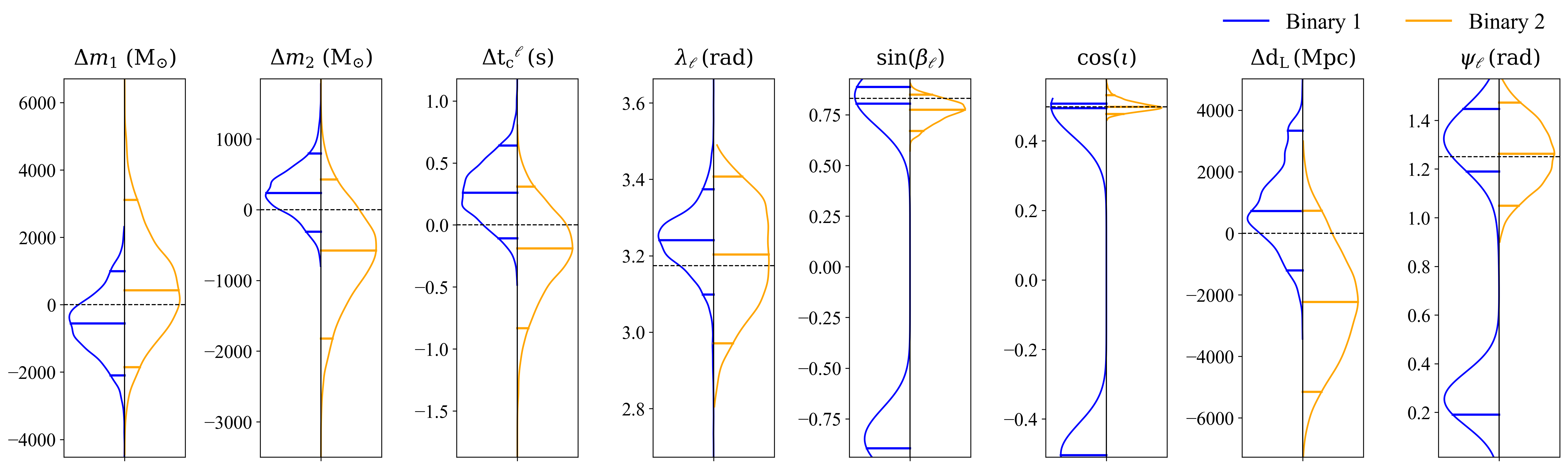}
\caption{Comparison of the marginalized posterior distributions obtained using 100 frequency bins for MBHBs 1 and 2 of Table \ref{tab:binary_params} injected with noise. The injected values of all the parameters for both binaries are recovered within 90\% CI. The posterior widths for component masses are broader for binary 2 as compared to binary 1 because of its shorter in-band duration and fewer number of cycles. The posteriors of binary 1 exhibit two latitudinal modes, whereas, the sampler picks only the correct mode in case of binary 2.}
\label{fig:binary1vsbinary2}
\end{figure*}

\begin{figure*}[t]
\centering
\includegraphics[width=\textwidth]{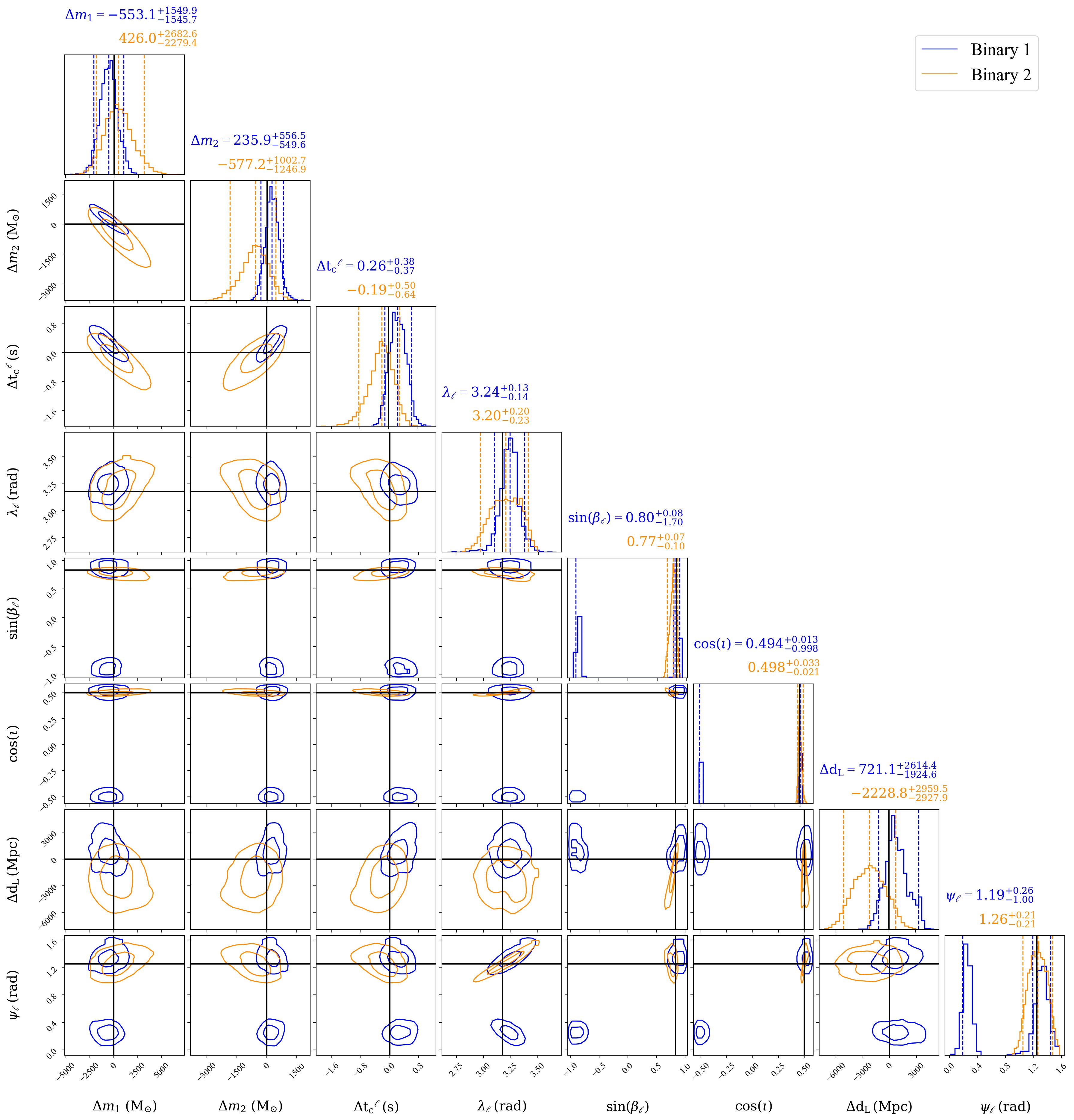}
\caption{Inferred parameter posterior distributions obtained using 100 frequency bins for MBHBs 1 and 2 of Table \ref{tab:binary_params} injected with noise. The contours in the marginalized 2-dimensional distributions corresponds to 50\% and 90\% CI and the black lines indicate the true parameter values. The labels at the top of each column indicate the median and 90\% CI for each of the parameters. The injected values of all the parameters for both binaries are recovered within the 90\% CI.}
\label{fig:binary1vsbinary2-corner}
\end{figure*}

We perform the meshfree parameter reconstruction using $N = 800$ randomly placed nodes centered around the injection point in the intrinsic parameter space composed of chirp mass and mass ratio. The boundary widths along chirp mass for binary 1 and 2 are $500 \msun$ and $2000 \msun$, respectively and the width along mass ratio for both binaries is $0.2$. The two widths along chirp mass and mass ratio for the two binaries result in almost similar widths in dimensionless chirp time coordinates. For robust inference of the binary parameters, the number of frequency bins to be considered should be adequate enough so that the piece-wise constant antenna pattern functions mimics the original functions. However, choosing a very large number of frequency bins will lead to more computational time. In order to find out the optimal choice for the number of frequency bins we carried out parameter estimation (PE) for MBHB 1 injected without noise realisation with 50, 75 and 100 frequency bins. The PE analyses were performed without noise so that the only source of errors could be attributed to the approximated antenna pattern functions. While evaluating the interpolated log-likelihood we only consider first $\ell = 15$ basis vectors for all the bins. For generating the RBF interpolants, we use a publicly available python package~\cite{RBF_github} and for sampling the posterior distribution we use the $\dynesty$ nested sampling package~\cite{Speagle:2019ivv} with $\mathtt{nlive} = 800, \: \mathtt{walks} = 150$, and $\mathtt{dlogZ} = 0.01$. With 50 frequency bins, several parameters including the component masses were not recovered within the 90\% credible interval (CI) suggesting that 50 bins are not sufficient to approximate the true antenna pattern functions. Fig.  \ref{fig:binary1-75vs100bins} shows the comparison of the marginalized posterior distributions of all eight parameters obtained using 75 and 100 frequency bins. The choice of using 75 frequency bins seems to recover all the parameters within 90\% CI but the inference gets improved further with 100 frequency bins. Note that the median values lie closer to the injected ones in the marginalized posteriors obtained with 100 frequency bins. Also, the marginalized posteriors of component masses are narrower with 100 frequency bins. The bimodality in the marginalized posteriors of $\beta_\ell$, $\iota$, and $\psi_\ell$ corresponds to the two latitudinal modes described above. The secondary mode in the marginalized posterior of $\psi_\ell$ appears close to $\mathtt{mod}\left(\psi_\ell + \frac{\pi}{2}; \frac{\pi}{2}\right)$ due to the chosen restricted prior range for $\psi_\ell$. The PE run employing 75 frequency bins took $\sim$562 CPU hours, as opposed to $\sim$750 CPU hours taken by the run with 100 bins. We carried out several PE analyses with noiseless injections similar to binary 1, but each injection differ in the sky location and coalescence time using 50, 75 and 100 bins. Similar trend was observed for all the injections in terms of frequency bins. All the parameters were accurately inferred using 100 frequency bins for all the injections. We also carried out analyses with even higher number of frequency bins, but adding more frequency bins does not improve the accuracy further. We note that the choice of 100 frequency bins is optimal in recovering posterior distribution. Furthermore, we performed parameter reconstruction of MBHBs 1 and 2 with stationary, Gaussian noise coloured with TDI-channel specific noise PSDs added to the injections. Figs. \ref{fig:binary1vsbinary2} and \ref{fig:binary1vsbinary2-corner} shows the comparison of the posterior distributions of all the parameters for the two binaries. All the parameters are recovered within the 90\% CI for both the  binaries. The width of the posteriors of component masses for the binary 2 is comparatively broader than binary 1 due to its relatively smaller in-band duration and fewer number of cycles. Also, the posteriors of binary 1 exhibits two latitudinal modes, whereas, the sampler picks only the right mode in case of binary 2. The PE run for binary 1 and 2 took $\sim$900 and $\sim$858 CPU hours, respectively. 
To demonstrate the accuracy of the RBF interpolation method we have taken posterior samples for binary 1~(see Table \ref{tab:binary_params}) in the intrinsic parameter space and have calculated the relative error between the RBF and the brute-force calculation of the likelihood function at these points. The median relative error of RBF interpolation is found to be $2.176 \times 10^{-5}$, indicating the good accuracy of RBF interpolation.
All the PE runs were performed on a single-core AMD EPYC 7542 2.90GHz CPU.

Similar runtimes have been obtained for heterodyned likelihood technique as implemented in $\bilby$~\cite{Hoy:2023ndx}, however, due to the difference in the considered parameter space and prior distributions in the analyses, an end-to-end comparison cannot be made. We would like to highlight that we have assumed no systematic bias in inference due to waveform mis-modeling, as same waveform approximant, $\imrpd$ is used for injection as well as recovery. Also, the point taken as the center of the parameter space region where interpolants are generated corresponds to the injected values of $\mchirp$ and $q$. The meshfree interpolants composed of RBFs have been found to be accurate only within a limited region in the intrinsic parameter space, which leads to restricted prior range over intrinsic parameters in the sampling stage. However, this limitation can be overcome easily by dividing the desired intrinsic parameter space into smaller patches and generating separate interpolants for each patch independently.

\section{Discussion and Conclusion}
\label{sec:discussion}
We have presented the extension of meshfree framework~\cite{Pathak:2022iar,Pathak:2023ixb} to infer the source properties of MBHBs that will be observed in LISA. The method was earlier applicable for parameter estimation of GW sources observed by second generation ground-based observatories. In this work, we performed inference on eight-dimensional parameter space from simulated data containing GW signal from a single MBHB. We show that properties of the LISA sources can be accurately inferred from GW signals in stationary, Gaussian instrumental noise using meshfree interpolation of the log-likelihood function.

Brute-force evaluation of the log-likelihood function is an expensive process due to two main computations---the evaluation of the waveform, and thereafter, evaluation of the overlap integrals appearing in Eq.~\eqref{eq:logl}. A sampler in a typical PE-run proposes $\mathcal{O}(10^6)$ points in the parameter space. Therefore, sampling the posterior distribution following the brute-force method becomes a great computational burden and leads to large runtime. The meshfree approximation of the log-likelihood function accelerates the computation by bypassing the two expensive computations during the sampling stage. The proposed technique can also be used for accelerating Bayesian reconstruction of CBC sources to be observed in the third-generation terrestrial detectors such as Cosmic Explorer~\cite{Reitze:2019iox,Hall:2020dps} and Einstein Telescope~\cite{Punturo:2010zz,Sathyaprakash:2012jk}.

We would like to mention that in this work, we have considered an idealized situation where a single GW signal is present in the LISA data. In actual scenario the data will contain multiple overlapping GW signals from all other source classes, not only the signals from MBHBs. Furthermore, the assumption of stationary and Gaussian instrument noise is also an ideal condition that we have restricted ourselves to. Due to data-gaps, glitches and expected change in sensitivity during the observation period, there will be non-stationary noise effects in the data. We have used approximant $\imrpd$ which is an aligned-spin waveform approximant and models only the dominant mode of GW emission from the binary. However, there will be contribution from higher order multipoles that $\imrpd$ does not account for. 

In this work, we have shown that the meshfree framework of accelerating the Bayesian parameter estimation of MBHBs, is a robust technique that leads to accurate inference of the source parameters. However, there are certain limitations of its current implementation that we highlight in this section. As mentioned above the restricted prior range over intrinsic parameters is used in our analyses during the sampling stage. The prior range is limited by the region considered for generating the meshfree interpolants of the desired quantities. We would like to overcome this issue by constructing separate interpolants for several non-overlapping patches in the intrinsic parameter space. Another scope of improvement is the strategy to place nodes in the intrinsic parameter space. Ideally, the nodes in the intrinsic parameter space should be placed in the region where likelihood has maximum support. Here, we have placed nodes uniformly inside a rectangular boundary around the injected parameters, a better strategy would be to place nodes inside a constant match contour. In this study, we restricted ourselves to spinless binaries and worked with only two-dimensional intrinsic parameter space composed of chirp mass and mass ratio of the binary. In future, we would like to extend our analysis for the aligned spin MBHBs. As described in Section \ref{subsec:Meshfree_likelihood_interpolation}, the evaluation of log-likelihood at a single point in parameter space requires calculating $\vec{z}^{(k)}$ and $\sigma^{2(k)}$ over several frequency bins. The computational cost of evaluating the log-likelihood scales linearly with the number of bins used to evaluate the integrals appearing in Eq.~\eqref{eq:logl2}. In this work, we have divided the full bandwidth into smaller frequency bins that are equally spaced on a logarithmic scale and found that 100 frequency bins are sufficient to approximate the antenna pattern functions. However, as evident from Figure \ref{fig:antenna-pattern}, the antenna pattern functions are almost constant for a wide range of frequencies and thus we do not need to place the bins uniformly. Moreover, most of the SNR contribution comes from the later part of the signal, the initial inspiral does not contribute significantly to the SNR, therefore, frequency bins can be constructed according to the SNR distribution across the spectrum. Thus, optimizing the number of frequency bins can greatly impact the speed-up gains of the log-likelihood function which we would like to explore in a future study.

\begin{acknowledgements}
A.~Sharma acknowledges useful discussions with Lalit Pathak throughout this work, Divya Tahelyani for helpful discussions and assistance with Mathematica, Sachin Shukla for helpful discussions during the initial phase of this work, and Shichao Wu for discussion on parameter space degeneracy. A.~Sharma also thanks IIT Gandhinagar for the senior research fellowship. A.~Sengupta thanks IIT Gandhinagar for the research support. A.~Sharma and A.~Sengupta thank TIFR Mumbai for the hospitality, where much of this work was conceptualized. 
The work of S.M. is a part of the $\langle \texttt{data|theory}\rangle$ \texttt{Universe-Lab}, supported by the TIFR  and the Department of Atomic Energy, Government of India. 
We acknowledge the computational resources provided by IIT Gandhinagar and the $\langle \texttt{data|theory}\rangle$ \texttt{Universe-Lab}. We also thank the high performance computing support staff at IIT Gandhinagar and TIFR for their help and cooperation.
\end{acknowledgements}
\section*{Data availability}
    The gravitational wave data that we analyze in this work is available in this repository~\cite{data_availability} together with the posterior samples of various parameter estimation analyses.
\appendix
\section{LISA response function}
\label{sec:lisa_response}
LISA will consist of three spacecrafts revolving about the Sun in an $\it{almost}$ equilateral triangular formation. The centroid of the triangular constellation will exhibit a circular orbit at $1 \mathrm{AU}$, trailing the Earth's orbit by $20^{\circ}$, with an orbital period of one year. In addition to the orbital motion, the entire apparatus will rotate about the centroid of the constellation in a clockwise direction as seen by an observer at the Sun. The rotational motion will have the same time-period of one year. The inter-spacecraft separation will be $L = 2.5 \times 10^6 \;\mathrm{km}$, and the changes in the separation will be tracked using laser ranging. Orbital motion of the observatory introduces amplitude and frequency modulation in the observed GW signal. The motion of the observatory results in a time dependent antenna response. Also, the wavelength of GW produced by MBHBs becomes comparable to the detector arms close to the merger. Therefore, finite size effects become important while calculating the antenna response function.


In this work, we adopt the rigid adiabatic approximation introduced in \cite{Rubbo:2003ap} and further used in \cite{Cornish:2020vtw} to model the antenna response of LISA. In this approximation, LISA is considered as a rigid triangle with fixed armlengths (ignoring the flexing of LISA arms) and the motion of the constellation is described as a sequence of stationary states. In each state, the detector is held motionless and the laser beam completes one round-trip across the LISA arms. This approximation holds when the timescale of frequency evolution of the chirping binary is longer compared to the light travel time between the arms. As shown below, the complete LISA response can be computed in the frequency domain by utilizing the time-frequency correspondence in stationary phase approximation, 
\begin{equation}
    t(f) = t_c - \frac{1}{2\pi}\frac{d\Psi(f)}{df},
    \label{eq:t-fmap}
\end{equation}
where, $\Psi(f)$ is the GW phase and $t_c$ is the coalescence time. For a given set of intrinsic parameters, at any given instantaneous frequency $f$ the corresponding $t$ can be evaluated using Eq.~\eqref{eq:t-fmap}. Thus, the time dependent quantities in the response function can be evaluated corresponding to a given frequency.

The complete antenna response is composed of single-link Doppler observables, i.e, the fractional frequency-shift of the laser as it travels along an arm of the detector. Adopting a coordinate system centered on the solar system barycenter (SSB frame), the fractional change in frequency of the laser due to the GW, when it travels from the sending spacecraft $s$ and arrive at the receiving spacecraft $r$ at the barycenter time $t$ is given by \cite{Creighton:2011zz},

\begin{equation}
y_{sr}(t)=\frac{\hat{\scriptr}_{s r} \otimes \hat{\scriptr}_{s r}}{2\left(1-\hat{k} \cdot \hat{\scriptr}_{s r}\right)}:\big[\mathbf{h}\left(t, \prescript{S}{}{\vec{\scriptr}}_r\right)-\mathbf{h}\left(t-L, \prescript{S}{}{\vec{\scriptr}}_s\right)\big],
    \label{eq:yij}
\end{equation}
where, $\hat{k}$ denotes the direction of the GW propagation, ${\hat{\scriptr}_{s r} \equiv \hat{\scriptr}_{s r}(t)}$ is a unit vector pointing from spacecraft $s$ to the spacecraft $r$ at time $t$, $\prescript{S}{}{\vec{\scriptr}}_s$ and $\prescript{S}{}{\vec{\scriptr}}_r$ are the positions of spacecrafts $s$ and $r$ with respect to SSB, respectively at time $t$. Note that in above expression we have taken ${c=1}$. And $\mathbf{h}$ is the transverse-traceless matrix representing the GW.
\begin{align}
    \nonumber
    \mathbf{h}(t, \prescript{S}{}{\vec{\scriptr}}) = \frac{1}{\dL}\big[A_+(\iota)\;& h_+(t, \prescript{S}{}{\vec{\scriptr}}, \intrinsic) \; \boldsymbol{\epsilon}^+ \\
    \label{eq:GW-general}
    &- A_\times(\iota)\; h_\times(t, \prescript{S}{}{\vec{\scriptr}}, \intrinsic) \; \boldsymbol{\epsilon}^\times\big],
\end{align}
where, $A_+(\iota) = (1 + \cos^2\iota)/2$ and $A_\times(\iota) = -\cos\iota$, $h_+$ and $h_\times$ are the plus and cross polarizations of the GW, respectively that depend only on the intrinsic parameters $\intrinsic$ of the source. The luminosity distance $\dL$ has been pulled out from $h_+$ and $h_\times$. For aligned-spin binaries, using the proportionality of two polarization states ${h_\times = -i h_+},$ Eq.~\eqref{eq:GW-general} can be written as,
\begin{equation}
    \mathbf{h}(t, \prescript{S}{}{\vec{\scriptr}}) = \frac{1}{\dL}\big[A_+(\iota)\; \boldsymbol{\epsilon}^+ + i A_\times(\iota)\; \boldsymbol{\epsilon}^\times\big] \; h_+(t, \prescript{S}{}{\vec{\scriptr}}, \intrinsic).
    \label{eq:GW}
\end{equation}
It will be useful to reference the spacecraft positions with respect to the center of constellation rather than the SSB. Thus, ${\mathbf{h}\left(t-L, \prescript{S}{}{\vec{\scriptr}}_s\right) = \mathbf{h}\left(t-L- \hat{k}\cdot\prescript{S}{}{\vec{\scriptr}}_0 - \hat{k}\cdot\vec{\scriptr}_s\right)}$ and ${\mathbf{h}\left(t, \prescript{S}{}{\vec{\scriptr}}_r\right) = \mathbf{h}\left(t - \hat{k}\cdot\prescript{S}{}{\vec{\scriptr}}_0 - \hat{k}\cdot\vec{\scriptr}_r\right)}$, where, $\prescript{S}{}{\vec{\scriptr}}_0$ is the position of center of constellation with respect to SSB center and $\vec{\scriptr}_s$ and $\vec{\scriptr}_r$ are the positions of spacecrafts $s$ and $r$ with respect to center of constellation, respectively. 

Upon taking Fourier transform of Eq.~\eqref{eq:yij} we get,
\begin{align}
    \nonumber
    \tilde{y}_{sr}(f) = &\frac{\hat{\scriptr}_{s r} \otimes \hat{\scriptr}_{s r}}{2\dL\left(1-\hat{k} \cdot \hat{\scriptr}_{s r}\right)}:\tilde{\mathbf{h}}(f)e^{-2 \pi i f (\hat{k}\cdot\prescript{S}{}{\vec{\scriptr}}_0)} \\
    &\quad\quad\big[e^{-2\pi i f L (\hat{k}\cdot\hat{\scriptr}_{sr} + \frac{\hat{k}\cdot\hat{\scriptr}_{s}}{\sqrt{3}})}
    - e^{-2\pi i f L (1 + \frac{\hat{k}\cdot\hat{\scriptr}_{s}}{\sqrt{3}})}\big],
    \label{eq:yij_fdomain}
\end{align}
where, we have substituted $\vec{\scriptr}_{r} = \vec{\scriptr}_{sr} + \vec{\scriptr}_s$ and used the fact that $\norm{\vec{\scriptr}_{s}} = L/\sqrt{3}$. Note that the time dependent vectors do not get affected by the Fourier transform because the detector is assumed to be motionless during the process. The phase shift of the laser corresponding to the change in frequency is obtained by integration, i.e., $\phi_{sr} = 2\pi \int y_{sr}(t)\; dt$. In frequency domain this phase shift can be written as,
\begin{align}
    \nonumber
    \tilde{\phi}_{sr}(f) = \frac{\pi L}{\dL} \left( \hat{\scriptr}_{sr} \otimes \hat{\scriptr}_{sr}\right):\tilde{\mathbf{h}}_0(f)\;&\mathrm{sinc}\bigg[\frac{f}{2f_\star}\left(1-\hat{k}\cdot\hat{\scriptr}_{sr}\right)\bigg] \\
    &e^{-i\frac{f}{2f_\star}\left(1 + \hat{k}\cdot\hat{\scriptr}_{sr} + \frac{2}{\sqrt{3}}\hat{k}\cdot\hat{\scriptr}_{s} \right)},
    \label{eq:phase_ij}
\end{align}
where, $\tilde{\mathbf{h}}_0(f)$ denotes the time-shifted GW as observed at the center of constellation, 
\begin{equation}
    \tilde{\mathbf{h}}_0(f) = \tilde{\mathbf{h}}(f)e^{-2 \pi i f (\hat{k}\cdot\prescript{S}{}{\vec{\scriptr}}_0)},
    \label{eq:h_LISA}
\end{equation}
and $f_\star = 1/2\pi L$ is the frequency corresponding to wavelength $L$ (transfer frequency). The total phase shift of laser as it travels from spacecraft $s$ to $r$ and arrive back to $s$ at the Barycenter time $t$ is given by,
\begin{equation}
    \tilde{\phi}_{\mathrm{total}}(f) = \pi L \: \tilde{\mathbf{h}}_{\mathrm{LISA}}(f),
    \label{eq:total_phi}
\end{equation}
where, $\tilde{\mathbf{h}}_{\mathrm{LISA}}(f)$ is the projected GW signal onto the LISA arm,
\begin{equation}
    \tilde{\mathbf{h}}_{\mathrm{LISA}}(f) = \frac{1}{\dL}\left( \hat{\scriptr}_{sr} \otimes \hat{\scriptr}_{sr}\right):\tilde{\mathbf{h}}_0(f)\; \mathcal{T}_{sr}(f),
    \label{eq:Det_frame_signal}
\end{equation}
where, the transfer function $\mathcal{T}_{sr}$ is given by,
\begin{align}
    \nonumber
    \mathcal{T}_{sr}(f, \hat{k}) = &\mathrm{sinc}\bigg[\frac{f}{2f_\star}\left(1-\hat{k}\cdot\hat{\scriptr}_{sr}\right)\bigg]e^{-i\frac{f}{2f_\star}\left(3 + \hat{k}\cdot\hat{\scriptr}_{sr} + \frac{2}{\sqrt{3}}\hat{k}\cdot\hat{\scriptr}_{s} \right)} \\
    &+ \mathrm{sinc}\bigg[\frac{f}{2f_\star}\left(1+ \hat{k}\cdot\hat{\scriptr}_{sr}\right)\bigg]e^{-i\frac{f}{2f_\star}\left(1 + \hat{k}\cdot\hat{\scriptr}_{sr} + \frac{2}{\sqrt{3}}\hat{k}\cdot\hat{\scriptr}_{s} \right)}.
    \label{eq:transfer_function}
\end{align}

The Michelson signal is constructed by calculating the difference in phase shifts induced across two different arms. For example, the Michelson signal measured at the spacecraft $1$ will be given by,
\begin{equation}
    s_1(t) \approx \big[\phi_{12}(t-L) + \phi_{21}(t) - (\phi_{13}(t-L) + \phi_{31}(t))\big].
    \label{eq:Michelson_signal}
\end{equation}
However, for a detector like LISA, the Michelson signal will be swamped by laser phase noise. Therefore, a technique called $\it{time}$--$\it{delay \;interferometry}$~\cite{Armstrong_1999,Tinto:2020fcc} is employed to synthesize the signals with suppressed laser phase noise. In this technique, several single arm round-trip phase shifts calculated at different delayed times are linearly combined to produce the TDI variables. One such variable is the X TDI channel composed of the following combination of two Michelson signals measured at the spacecraft 1:
\begin{equation}
    X(t) = s_1(t) - s_1(t - 2L).
    \label{eq:X-TDI}
\end{equation}
Similarly, the Y and Z TDI channels can be constructed from the Michelson signals measured at spacecraft 2 and 3, respectively. The Fourier transform of Eq.~\eqref{eq:X-TDI} gives the frequency domain X TDI channel, as, 
\begin{equation}
    \tilde{X}(f) = 2 i e^{-i\frac{f}{f_\star}} \sin(\frac{f}{f_\star})\tilde{s}_1(f)
\end{equation}

\begin{align}
    \nonumber
    \tilde{X}(f) = &- \frac{1}{\dL}\frac{f}{f_\star} \; e^{-i\frac{f}{f_\star}} \sin(\frac{f}{f_\star})
        \left[ \left( \hat{r}_{12} \otimes \hat{r}_{12}\right)\mathcal{T}_{12}(f) \right . \\
    & \quad \left . - \left( \hat{r}_{13} \otimes \hat{r}_{13} \right)\mathcal{T}_{13}(f) \right] :\tilde{\mathbf{h}}_0(f).
\end{align}
Note that the factor $-f/f_\star$ is multiplied to be consistent with fractional-frequency TDI response as used by LDC data sets. We have also suppressed the phase factor of $\pi/2$. If we define $d_{sr}^{+, \times} = (\hat{\scriptr}_{sr} \otimes \hat{\scriptr}_{sr}) : \boldsymbol{\epsilon}^{+, \times}$ and use Eq.~\eqref{eq:GW}, the above expression can be written as,
\begin{align}
    \nonumber
    \tilde{X}(f) = \frac{1}{\dL}\big[F_X^+(f, \vec{\alpha}) &A_+ + i F_X^\times(f, \vec{\alpha}) A_\times\big] \\
    &e^{-2 \pi i f (\hat{k}\cdot\prescript{S}{}{\vec{\scriptr}}_0)}\tilde{h}^{\mathrm{TDI}}_+(f),
    \label{eq:X-TDI_complete}
\end{align}
where, $\tilde{h}^{\mathrm{TDI}}_+(f)$ is given by the product of TDI induced pre-factor and $\tilde{h}_+(f)$,
\begin{equation}
    \tilde{h}^{\mathrm{TDI}}_+(f) = -\frac{f}{f_\star} e^{-i\frac{f}{f_\star}} \sin(\frac{f}{f_\star})\tilde{h}_+(f),
    \label{eq:h_tdi}
\end{equation}
and, $F_X^{+, \times}$ are given by,
\begin{equation}
    F_X^{+, \times}(f, \vec{\alpha}) = [d_{12}^{+, \times} \mathcal{T}_{12}(f, \hat{k}) -d_{13}^{+, \times}\mathcal{T}_{13}(f, \hat{k})],
    \label{eq:G_X}
\end{equation}
and, $\vec{\alpha} = (\lambda, \beta, \psi)$ denotes the extrinsic parameters, namely ecliptic longitude ($\lambda$), ecliptic latitude ($\beta$) and polarization angle ($\psi$). Note that in Eq.~\eqref{eq:h_tdi} TDI induced pre-factor is completely frequency dependent, therefore, we can club it with $\tilde{h}_+(f)$ and denote $\tilde{h}^{\mathrm{TDI}}_+(f)$ as $\tilde{h}_+(f)$ omitting superscript $\mathrm{TDI}$. Similar expressions for the antenna pattern functions as Eq.~\eqref{eq:G_X} can be obtained for the Y and Z TDI channels by a cyclic permutation of the indices $(1, 2, 3)$ in the above expression. However, the noise is correlated in the X, Y and Z TDI channels, noise orthogonal TDI channels A, E and T (named after Armstrong, Estabrook and Tinto; the pioneers of the time-delay interferometry technique.) can be obtained from the linear combinations of X, Y and Z channels and are given by,
\begin{align}
    \tilde{A} = &\frac{1}{\sqrt{2}}(\tilde{Z} - \tilde{X}) \\
    \tilde{E} = &\frac{1}{\sqrt{6}}(\tilde{X} - 2\tilde{Y} + \tilde{Z}) \\
    \tilde{T} = &\frac{1}{\sqrt{3}}(\tilde{X} + \tilde{Y} + \tilde{Z}).
    \label{eq:AETchannels}
\end{align}
In this study, we generate the simulated data in the TDI channels A, E and T.

\begin{widetext}
    \section{Noise Power Spectral Density (PSDs)}
    \label{sec:noise-psds}
    
    The PSDs used for various TDI channels are obtained from~\cite{Sangria_LDC,Babak:2021mhe} and are given by,
    
    \begin{equation}
        S_n^A(f) = S_n^E(f) = 8 \sin ^2 \big(\frac{f}{f_*}\big)\bigg[4\bigg(1+\cos\big(\frac{f}{f_*}\big)
        +\cos ^2 \big(\frac{f}{f_*}\big)\bigg) S^{A c c}(f)
        +\bigg(2+\cos \big(\frac{f}{f_*}\big)\bigg) S^{I M S}(f)\bigg],
    \end{equation}
    
    \begin{equation}
        S_n^T(f) = 32 \sin ^2 \big(\frac{f}{f_*}\big) \sin ^2 \big(\frac{f}{2f_*}\big)\bigg[4 \sin ^2 \big(\frac{f}{2f_*}\big) S^{A c c}(f)
        +S^{I M S}(f)\bigg],
        \label{eq:psd-T}
    \end{equation}
    where, $S^{A c c}(f)$ and $S^{I M S}(f)$ corresponds to the contributions from acceleration noise and interferometric measurement system noise, respectively, and are given by,
    \begin{equation}
        S^{A c c}(f) = \bigg(\frac{3\times10^{-15}}{2\pi fc}\bigg)^2 \bigg[1 + \bigg(\frac{4\times10^{-4}}{f}\bigg)^2\bigg]
        \times \bigg[1 +\bigg(\frac{f}{8\times10^{-3}}\bigg)^4\bigg],
    \end{equation}
    
    \begin{equation}
        S^{I M S}(f) = 2.25\times10^{-22}\bigg(\frac{2\pi f}{c}\bigg)^2\bigg[1 + \bigg(\frac{2\times10^{-3}}{f}\bigg)^4\bigg].
    \end{equation}
\end{widetext}

\bibliography{references}

\end{document}